\documentclass[reprint,amssymb,aps,pre,superscriptaddress,nobibnotes,
]{revtex4-2}
\usepackage{soul,color,xcolor}
\usepackage{amsmath,amsthm}
\usepackage{array}
\usepackage{subfigure}
\usepackage{graphicx}
\usepackage{bm}
\usepackage[colorlinks,
          linkcolor=black,
            citecolor=black,
            urlcolor=blue
           ]{hyperref}
\usepackage{comment}
\usepackage{mathrsfs}
\usepackage{physics}

\soulregister\cite7
\soulregister\ref7 
\soulregister\citep7 
\soulregister\citet7 
\soulregister\pageref7 

\begin{document}
\title{Entangled biphoton generation in myelin sheath}
\author{Zefei Liu}
\email{1239463505@shu.edu.cn}
\author{Yong-Cong Chen}
\email[Corresponding author: ]{chenyongcong@shu.edu.cn}
\affiliation{Shanghai Center for Quantitative Life Sciences \& Physics Department, Shanghai University, Shanghai 200444, China}
\author{Ping Ao}
\affiliation{College of Biomedical Engineering, Sichuan University, Sichuan 610065, China}
\begin{abstract}
  Consciousness within the brain hinges on the synchronized activities of millions of neurons, but the mechanism responsible for orchestrating such synchronization remains elusive. In this study, we employ cavity quantum electrodynamics (cQED) to explore entangled biphoton generation through cascade emission in the vibration spectrum of C-H bonds within the lipid molecules' tails. The results indicate that the cylindrical cavity formed by a myelin sheath can facilitate spontaneous photon emission from the vibrational modes and generate a significant number of entangled photon pairs. The abundance of C-H bond vibration units in neurons can therefore serve as a source of quantum entanglement resources for the nervous system. The finding may offer insight into the brain's ability to leverage these resources for quantum information transfer, thereby elucidating a potential source for the synchronized activity of neurons. 
\end{abstract}
\maketitle


\section{Introduction}

Understanding the intricacies of the human brain and its functions has perennially posed an intriguing and challenging puzzle. The synchronization of neurons within the cerebral cortex serves as the foundation for diverse neurobiological processes\cite{longrangecommunicate}, closely linked to anomalies in brain function and brain diseases\cite{reviewneuralsynchrony}. Notably, Parkinson's disease manifests a loss of neural activity synchronization in regions with damaged neurons\cite{wang}. Despite these observations, the mechanisms underpinning precise synchronization of neural activity remain unknown\cite{HAMEROFF201439}, necessitating interdisciplinary research, particularly in the realms of neuroscience and quantum physics.

In recent decades, quantum computing, harnessing the unique features of quantum entanglement, has witnessed remarkable success\cite{naturequcomp}. Experiments validating nonlocal correlations in quantum entanglement\cite{RevModPhys.81.865} have enabled quantum computation to outpace classical counterparts in tasks such as the Shor\cite{shor} and Grover\cite{grover} algorithms. Quantum computing's applicability to neuroscience was initially proposed by Penrose, who suggested a role for microtubules in quantum computation within the brain\cite{HAMEROFF201439}, and further explored for example,  by Fisher who proposed nuclear spins as mediators\cite{FISHER2015593}. Despite experimental deviations from predictions of these models\cite{penrosefa,fisherfa}, the nonlocal correlations inherent in quantum entanglement remain captivating.

Recent studies highlight the role of photon as a quantum object not only in plants and bacteria but also in animal life activities\cite{futureqb}. Examples include mid-infrared (MIR) photons from ATP hydrolysis driving DNA replication\cite{atp} and polaritons formed by visible light  coherently and resonantly coupled to the excitons of chlorophyll molecules in chloroplasts facilitating efficient energy transfer in photosynthesis\cite{chenycprl}. Ultra-weak photon emission (UPE) in living organisms, traditionally considered as metabolic by-products, is now implicated in neuronal function\cite{ultraweak}. Moreover, MIR photons at 53.7 THz modulate K$^+$ ion channel activity, neuronal signaling, and sensorimotor behavior\cite{songpnas}. Photons released during the TCA cycle in neurons resonantly couple to C-H bond vibrons in lipid molecules, altering possibly the dielectric constant of the membrane to enhance action potential conduction\cite{songneuron}. These findings, though each may require further scrutiny,  offer an alternative perspective on the significance of light in neural activity.

Myelin sheath, a lipid membrane encasing the outer side of a neuron's axon, provides energy to axon, enhances action potential conduction, and acts as an insulator in the nervous system\cite{axonstruc}. Abnormal myelin function or damage to the myelin structure is strongly associated with neurodegenerative diseases such as multiple sclerosis and Alzheimer's disease\cite{myelindisease}. Myelin sheath is generally regarded only as an insulator. However, emerging evidence suggests myelin's plasticity, indicating its role beyond insulation and its potential to promote neural phase synchronization\cite{myelinplastic}.

This study demonstrates that the vibrational spectrum of C-H bonds in lipid molecule's tails, within cylindrical cavities formed by myelin sheath encasement, can generate quantum entangled photon pairs through cascade radiation from the second excited state to the ground state. In Sec.\ref{sec:cylindrical}, we establish the axon's well-defined cylindrical structure under the myelin sheath and discuss the quantization of the electromagnetic field within the cylindrical cavity. We show that, within the infrared region and under the dipole approximation, two-photon processes in the vibrational spectrum are predominantly governed by cascade radiation from dipole interactions in Sec.\ref{sec:cascade}. Utilizing Schmidt analysis, we assess the degree of quantum entanglement in biphoton systems and exemplify the potential for generating quantum entanglement in neural systems using real structural data of myelinated neuron from experiments in Sec.\ref{sec:schmidt}. Leveraging the nonlocal correlation properties of quantum entanglement, one may speculate that quantum entanglement will effectively synchronize neuronal activity throughout the brain, shedding light on the 'synchronization' puzzle in consciousness.

\section{Cylindrical cavity formed by myelin sheath}\label{sec:cylindrical}

\begin{figure*}[ht]
    \includegraphics[scale=0.4]{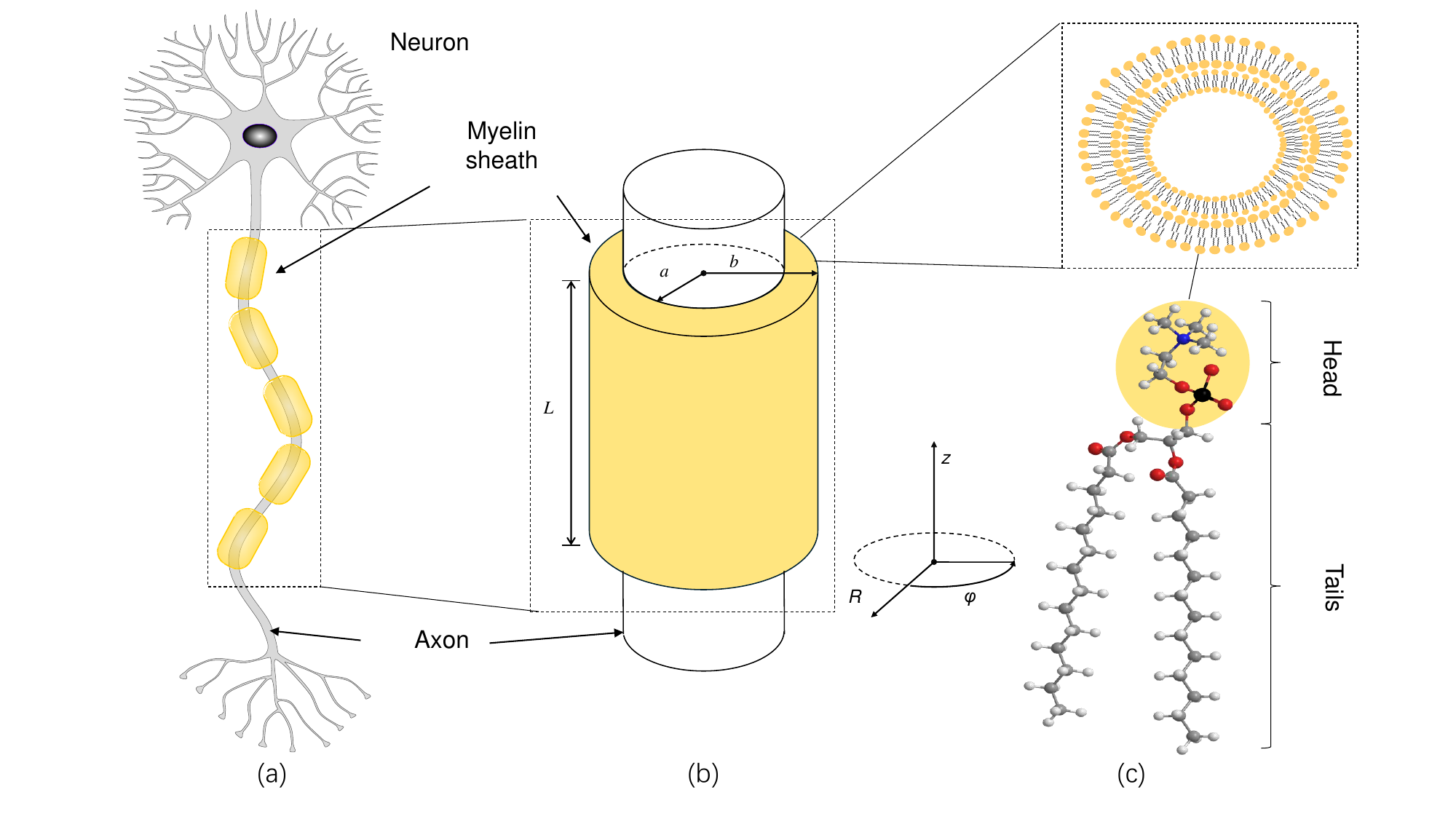}
    \caption{Cylindrical cavity formed by myelin sheath. (a) Axons of myelinated neuron have several segments of myelin sheaths wrapped in lipid membranes of different oligodendrocytes. The gap between the two segments of the myelin sheath is called node of Ranvier and is usually 1$\sim$2 $\mu$m in length, whereas each segment of the myelin sheath is typically around 100 $\mu$m in length. The gap in the node of Ranvier is negligible for the whole axon. (b) Consider the entire myelin-coated axon as a cylinder of length $L$, with the radius of the bare axon as $a$, the radius of the coated as $b$, and the central axis as the z-axis. (c) Phospholipid molecules, as the major component of myelin, have tails consisting of a large number of C-H bonds.}
    \label{fig:myelincylinder}
\end{figure*}

As shown in Figure.\ref{fig:myelincylinder}(a), oligodendrocytes in the central nervous system (CNS) form myelin sheaths by wrapping lipid-rich membranes around axons, and because of their high content of myelin, which is white in colour, axons surrounded by myelin sheaths in the CNS are called white matter\cite{shaotu2myelin}. More than half of the human brain is white matter, which supports the rapid and simultaneous transmission of information between numerous grey matter areas of the CNS\cite{myelinplastic}. The formation of microcavities is crucial in exciton-polaron mediated energy transfer in the recent progress of explaining the mechanism of high efficient energy transfer\cite{chenycprl,Luwenbin}. The total reflection of light satisfying the evanescent wave condition on the inner wall of the cavity causes the confinement of photons by the cavity. The myelin sheath, which generally consists of 10$^2$ lipid bilayers (Figure.\ref{fig:myelincylinder}(c)) wrapped around the axon, serves mainly as an insulator in neurons and can have a better confinement effect on photons due to the formation of polaritons in the myelin sheath\cite{songneuron}. For simplicity and without loss of generality, it is reasonable to consider the outer wall of the myelin sheath as a perfect conductor wall, considering only that electromagnetic fields do not leak outside the myelin sheath. Concentrically wrapped around the periphery of the axon, myelin together with the axon forms a cylinder-like structure, and based on this fact, we consider myelinated neuronal axons as cylinders (Figure.\ref{fig:myelincylinder}(b)). The thickness of the myelin sheath is generally in the range of 1$\sim$3 $\mu$m \cite{AxonPhysiology}. The ratio of the length between two neighbouring nodes of Ranvier to the thickness of the myelin sheath is $\sim$100 , and the width of the nodes of Ranvier is generally in the range of 1$\sim$2 $\mu$m\cite{ratio10078}, so that we can ignore the gaps caused by the nodes of Ranvier, and consider the multiple myelin sheaths merged together as a whole. In this way, we view the myelin-coated axonal portion of the entire myelinated neuron as a conductor-walled cylinder, as shown in Figure.\ref{fig:myelincylinder}(b). The C-H bond dipole in the lipid tail lies in between the inner and the outer radii $a$ and $b$.

The Hamiltonian of the quantised electromagnetic fields and the electric field inside the cylindrical cavity can be written respectively as\cite{kakazucylinder}
\begin{align}
  \bm{E}(\bm{R},t)&=\sum_{s\sigma}f_{s\sigma}(a_{s\sigma}(t)+a^{\dagger}_{s\sigma}(t))\bm{u}_{s\sigma}(\bm{R}) ,\\
  H_R&=\sum_{s\sigma}\ \hbar\omega_{s\sigma}\left(a_{s\sigma}^\dag a_{s\sigma}+\frac{1}{2}\right) ,
\end{align}
where $\bm{u}_{s\sigma}$ and $a_{s\sigma}$ are respectively the vector mode functions and annihilation operator and $f_{\omega}=\sqrt{\hbar\omega_{s\sigma}/({2\epsilon_0})}$. $s=(m, \mu ,n)$ is the mode index and $\sigma=0,1$ represents polarization.

The vector mode functions $\bm{u}_{s\sigma}(\sigma=0,1)$ take the form
\begin{align}
  \vb*{u}_{s1}=\nabla\times\nabla\times\vb*{e_z}\psi_{s1}\label{us1} ,\\
  \vb*{u}_{s2}=i\omega_{s2}\nabla\times\vb*{e_z}\psi_{s2}\label{us2} ,
\end{align}
where $\psi_{s1}$ and $\psi_{s2}$ have the form
\begin{align}
  \psi_{s1}=c_{s1}J_m(\frac{\chi_{s1}R}{b})e^{im\varphi}\cos{(\frac{n\pi z}{L})} ,\\
  \psi_{s2}=c_{s2}J_m(\frac{\chi_{s2}R}{b})e^{im\varphi}sin{(\frac{n\pi z}{L})} .
\end{align}
$\chi_{s1}\equiv\chi_{m\mu 1}$ is the $\mu$th zero point of the first kind of Bessel function $J_m$, $\chi_{s2}\equiv\chi_{m\mu 2 }$ is the $\mu$th zero of ${J^\prime}_m$(the derivative of $J_m$). $c_{s\sigma}$ is a normalization constant, $c_{s1}=\sqrt{({2c^2 b^2})/({V\alpha_{s1}\chi^2_{s1}\omega^ 2_{s1}})}$, $c_{s2}=\sqrt{({2 b^2})/({V\alpha_{s2}\chi^2_{s2}\omega^2_{s2}})}$, Here $\alpha_{s1}\equiv\alpha_{m\mu 1}=J^2_{m+1}(\chi_{m\mu 1})$, $\alpha_{s2}\equiv\alpha_{m \mu 2}=J^2_m(\chi_{m\mu 2})-J^2_{m+1}(\chi_{m\mu 2})$. $L$ is the length of the cylindrical cavity. The eigen modes of the cavity are determined by
\begin{equation}
    \omega_{s\sigma} = c\sqrt{\left(\frac{\chi_{s\sigma}}{b}\right)^2+\left(\frac{n\pi}{L}\right)^2} .
\end{equation}

\section{radiative cascade within three-level morse oscillator}\label{sec:cascade}
In this section, we first introduce morse oscillator which is extensively used for depicting the vibration of chemical bonds. Then we discuss two-photon processes via radiative cascade by dipole interaction. For obtaining actual wave function of biphoton, we use the real size data of myelinated neuron by experiments.
\subsection{Chemical Bonds Described by Morse Oscillators}
\begin{figure*}[ht]
    \centering
    \includegraphics[width = 17.2cm]{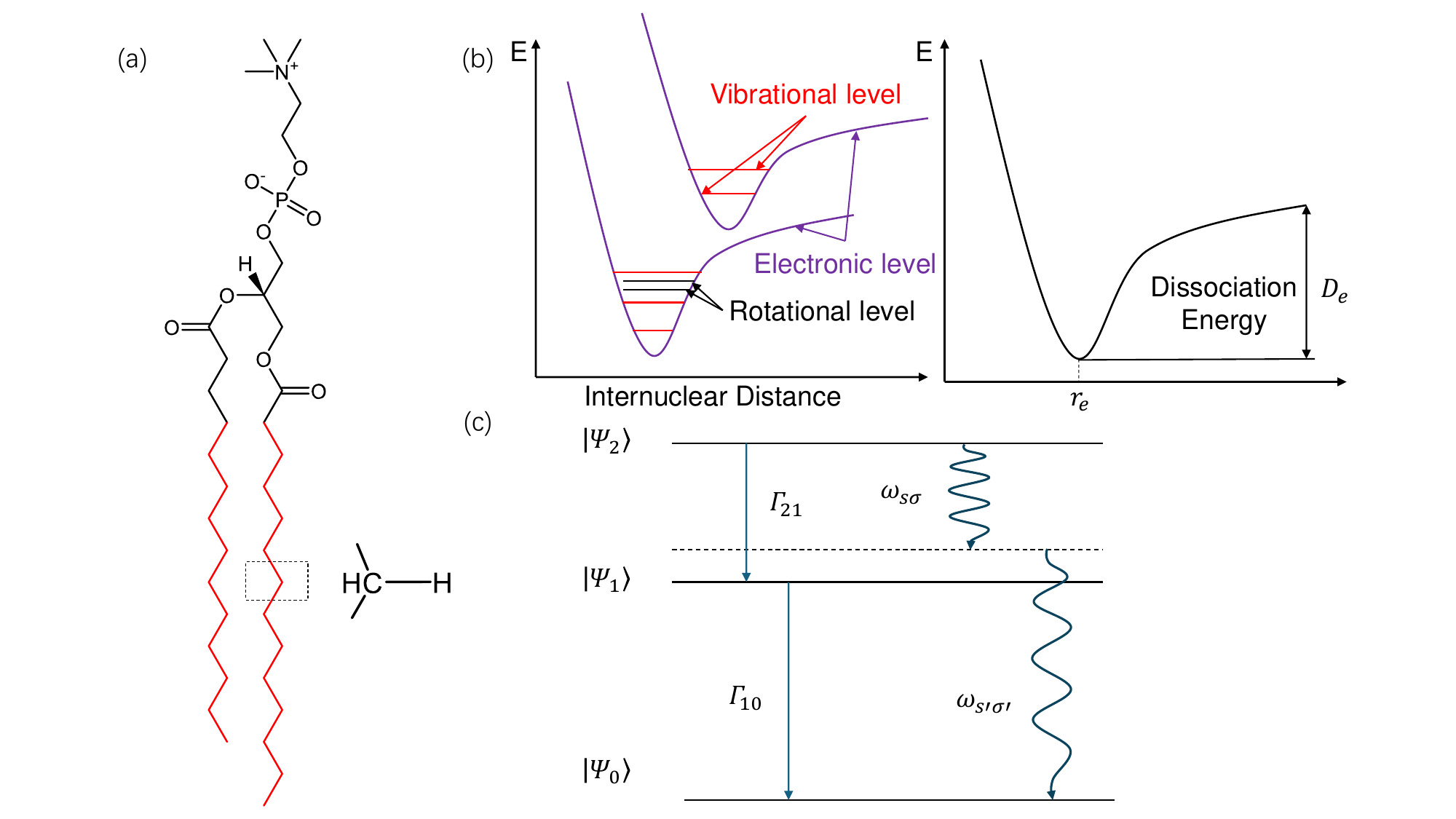}
    \caption{Cascade emission from vibrational spectrum. (a) A skeletal formula of a phospholipid molecule, with the red part representing the carbon chain in the tail, consisting of several methylene groups and two methyl groups. (b) Left: Schematic diagram of molecular energy levels, with electronic energy levels indicated in purple, rotational energy levels in black, and vibrational energy levels in red. Right: Vibrational energy levels are generally represented by Morse oscillators. The lowest point of the potential energy corresponds to the equilibrium bond length of the chemical bond $r_e$, and the depth of the potential energy characterizes the dissociation energy $D_e$. (c) Cascade emission in three-level system. This figure represents a three-level system returning to the ground state from the second excited state by emitting photons with frequencies $\omega_{s\sigma}$ and $\omega_{s'\sigma'}$. $\varGamma_{21}$ and $\varGamma_{10}$ are the transition rates(or the reciprocal of lifetime) between states.}
    \label{fig:cascade}
\end{figure*}

In contrast to the rotation of a chemical bond and its electronic energy levels (Figure.\ref{fig:cascade}(b)), the transitions in the vibrational energy spectrum are typically on the order of 0.1 eV, placing them in the mid-infrared band. This band lies in the range that has a significant influence on neural activity. The vibrations of chemical bonds are generally anharmonic oscillators, and the energy levels are not equally spaced. In order to describe anharmonic vibrations, we adopt Morse oscillators\cite{morse} to describe actual chemical bond vibrations as shown in Figure.\ref{fig:cascade}(b). The advantage of using a Morse oscillator is that its determinant Schrödinger equation has an analytical solution, which makes it very convenient for us to calculate the dipole moment of the chemical
bond. the potential function of Morse oscillator can be written as\cite{morsepotential}
\begin{equation}
    V\left(r\right)=D_e\left(e^{-2w\left(r-r_e\right)}-2e^{-w\left(r-r_e\right)}\right) .
\end{equation}
where $r_e$ is bond length at equilibrium. $D_e$ and $w$ represent the depth and the width of the potential well respectively. Its stationary Schrödinger equation $H_P\ket{\varPsi_\nu}=E_{\nu}\ket{\varPsi_\nu}$ has an analytical solution, which is
\begin{align}
    &\ket{\varPsi_\nu}=N_\nu z^{\lambda-\nu-\frac{1}{2}}e^{-\frac{1}{2}z}L_\nu^{\left(2\lambda-2\nu-1\right)}\left(z\right) ,\\
    &E_\nu^\prime=-\left(\lambda-\nu-\frac{1}{2}\right)^2 \\
    &\nu=0,1,\cdots, [N] =[\lambda-\frac{1}{2}] \notag ,
\end{align}
with variable substitutions $z=2\lambda e^{-w\left(r-r_e\right)}$ and $E_\nu^\prime=[({2m})/({w^2\hbar^2})]E_\nu$. In the above, $L_\nu^{\left(2\lambda-2\nu-1\right)}\left(z\right)$ is a generalized Laguerre polynomial, $[ \lambda-\frac{1}{ 2}]$ represents the largest integer not exceeding $[ \lambda-\frac{1}{2}]$, and the normalization constant $N_\nu=\sqrt{[{n!(2\lambda-2 \nu-1)}]/[{\mathrm{\Gamma}\left(2\lambda-\nu\right)}]}$ with $\mathrm{\Gamma} \left(x\right)$ being the gamma function.

The matrix elements of positional operators also have the following analytical forms\cite{matrixmorse}
\begin{equation}
\begin{split}
      \mel{\varPsi_\nu}{r}{\varPsi_{\nu'}}=&\frac{2\left(-1\right)^{\nu^\prime-\nu+1}}{\left(\nu^\prime-\nu\right)\left(2N-\nu-\nu^\prime\right)}\\
      &\times\sqrt{\frac{\left(N-\nu\right)\left(N-\nu^\prime\right)\mathrm{\Gamma}\left(2N-\nu^\prime+1\right)\nu^\prime!}{\mathrm{\Gamma}\left(2N-\nu+1\right)\nu!}} .\label{matele}
\end{split}
\end{equation}

\subsection{Biphoton Wave Function via Cascade Radiation}
The Hamiltonian of the interaction of matter with an electromagnetic field is usually written as\cite{cohen1998atom}
\begin{equation}
  H_I = -\frac{e}{m} \bm{p}\cdot \bm{A} + \frac{e^2}{2m} \bm{A}^2 + \hbar g^s \bm{s}\cdot\bm{B} ,
\end{equation}
where $\bm{p}$ and $\bm{A}$ are the momentum of the matter and vector potential of electromagnetic field, respectively, and $g^s$ is the gyromagnetic ratio of the particle. The last two terms describe two-photon processes under strong electromagnetic fields and strong interactions\cite{cohen1998atom}, respectively, which should be ignored considering that strong field environments are not usually existed in neurons. Under the long-wave approximation (or dipole approximation), the first term can be written as $-\bm{D}\cdot\bm{E}$, where $\bm{D}$ and $\bm{E}$ are the dipole moment of the particle and electric field respectively. The process of emitting two photons by jumping from the first excited state to the ground state under dipole interaction is forbidden, so we will consider the process of emitting two photons from the second excited state to the ground state by cascade radiation\cite{scully1999quantum}, as shown in Figure.\ref{fig:cascade}(c).

As the vibrational displacement of the chemical bond is usually in the order of 10$^{-1}$ \AA, the orders of magnitude of the coupling constants $g_{\omega}$ and $g_{\omega^\prime}$ below are about 10$^7$ Hz. This is much smaller than $\sim 10^{14}$ Hz from the vibrational levels themselves. Therefore a weak coupling condition is generally satisfied and we can consider the so-called rotational wave approximation (RWA) under which the  quickly oscillating terms in the Dirac interacting picture are dropped. Keeping in mind though there could be problems with RWA as the coupling gets strong\cite{Fleming_2010rwa}. The approximation can serve as a starting point in the current context for the cascade emissions of interest arise mainly from slow oscillating processes that conserve the total energy.

With RWA, the interacting Hamiltonian under the Dirac picture can be written as\cite{scully1999quantum}
\begin{equation}
  \begin{split}
    H_I(t)=&\hbar\sum_{s\sigma}\left[g_\omega^\ast\sigma_+^{\left(1\right)}a_{s\sigma}e^{i\left(\omega_{21}-\omega_{s\sigma}\right)t}+h.c.\right]\\
      &+\hbar\sum_{s^\prime\sigma^\prime}\left[g_{\omega^\prime}^\ast\sigma_+^{\left(2\right)}a_{s^\prime\sigma^\prime}e^{i\left(\omega_{10}-\omega_{s\sigma}\right)t}+h.c.\right] ,
\end{split}
\end{equation}
where $\sigma_+^{\left(1\right)}=\ket{\varPsi_2}\bra{\varPsi_1}$, $\sigma_+^{\left(2 \right)}=\ket{\varPsi_1}\bra{\varPsi_0}$, and $g_\omega$ and $g_{\omega\prime}$ are states $\ket{\varPsi_2}$ to state $\ket{\varPsi_1}$ and state $\ket{\varPsi_1}$ to $\ket{\varPsi_0} $ state transition coupling constant which take the form of
\begin{align}
  g_{\omega}=\frac{f_{\omega}\mel{\varPsi_1}{\bm{D}}{\varPsi_2}\cdot\bm{u}_{s\sigma}}{\hbar} ,\\
  g_{\omega'}=\frac{f_{\omega'}\mel{\varPsi_0}{\bm{D}}{\varPsi_1}\cdot\bm{u}_{s'\sigma'}}{\hbar} .
\end{align}

Similarly to the Jaynes-Cummings model\cite{Cummings_2013}, we let the total number of excitations between the atom and the cavity mode be conserved, and let the vibrational energy levels and the state vectors of the photons be of the form
\begin{equation}\label{matterfockstate}
\begin{split}
\ket{\Psi(t)}=c_0(t)\ket{\varPsi_2,0}&+c_1(t)\sum_{s\sigma}\ket{\varPsi_1,\omega_{s\sigma}}\\
&+c_2(t)\sum_{s'\sigma'}\ket{\varPsi_0,\omega_{s\sigma},\omega_{s'\sigma'}} .
\end{split}
\end{equation}

Substituting Eq.(\ref{matterfockstate}) into the time-dependent Schr\"{o}dinger equation, we have three equations of motion for the probability amplitudes
\begin{align}
  {\dot{c}}_0&=-i\sum_{s\sigma} g_\omega^\ast c_1e^{i\left(\omega_{21}-\omega_{s\sigma}\right)t}\label{c0dot} ,\\
  {\dot{c}}_1&=-ig_\omega c_0e^{i\left(\omega_{21}-\omega_{s\sigma}\right)t}-i\sum_{s^\prime\sigma^\prime} g_{\omega^\prime}^\ast c_2e^{i\left(\omega_{10}-\omega_{s^\prime\sigma^\prime}\right)t}\label{c1dot} ,\\
  {\dot{c}}_2&=-ig_{\omega^\prime}c_1e^{i\left(\omega_{10}-\omega_{s^\prime\sigma^\prime}\right)t}\label{c2dot} .
\end{align}

We begin by looking at the first two equations, and in order to simplify the system of equations, we first integrate Eq.(\ref{c2dot}) and substitute into Eq.(\ref{c1dot}) without considering $c_0$ to get
\begin{equation}
    {\dot{c}}_1 = -\sum_{s^\prime\sigma^\prime} \abs{g_{\omega^\prime}}^2 \int^t_0 \dd t' e^{i\left(\omega_{10}-\omega_{s^\prime\sigma^\prime}\right)(t-t')}c_1(t') .\label{c1dotint}
\end{equation}
This is still an exact equation. Next we make some approximations. The dipole moment of the C-H bond in the tail of the lipid molecule is overwhelmingly perpendicular to the radial direction, and for the sake of computational convenience, we assume that the dipole moment is only in the z-direction. It is obvious that $\vb*{u}_{s2}$ has no z component from Eq.(\ref{us2}), and we only need to find out the z component of $\vb*{u}_{s1}$. By expanding Eq.(\ref{us1}) we get $\vb*{u}_{s1} = \nabla(\nabla\cdot \vb*{e_z}\psi_{s1})-\laplacian\vb*{e_z}\psi_{s1}=k^2_{s1}\vb*{e_z}\psi_{s1}+\nabla(\partial_z \psi_{s1})$. Then we have
\begin{equation}
  \vb*{u}_{s1,z}=\vb*{e_z}\sqrt{\frac{2c^2\chi^2_{s1}}{Vb^2\alpha_{s1}\omega^2_{s1}}}J_m(\frac{\chi_{s1}R}{b})\cos{(\frac{n\pi z}{L})} e^{im\varphi}.
\end{equation}
Now we have explicit form of the coupling constant
\begin{equation}
    g_{\omega'} = d_{01}\sqrt{\frac{c^2\chi^2_{s'1}}{\hbar\epsilon_0 V b^2\alpha_{s1}\omega_{s'1}}}J_{m'}(\frac{\chi_{s'1}R}{b})\cos{(\frac{n'\pi z}{L})} e^{im\varphi} ,\label{couple}
\end{equation}
where $d_{01}$ is the matrix element of the dipole moment $\bm{D}$ which is $\mel{\varPsi_0}{\vb*{D}}{\varPsi_1}$.

In general the length of the myelin sheath is much larger than the radius of the axon, at which point the energy level gap of the cavity mode is very small, and we can convert the summation into integral in Eq.(\ref{c1dotint}), in other words
\begin{equation}
    \lim_{L\to\infty}\frac{1}{L}\sum_{n'} = \frac{1}{c\pi} \int_{c\chi_{s'1}/b}^{\infty}\frac{\dd \omega_{s'1} \omega_{s'1}}{\sqrt{\omega_{s'1}^2 - c^2\chi_{s'1}^2/b^2}} .\label{sum2int}
\end{equation}
Now Eq.(\ref{c1dotint}) becomes
\begin{widetext}
    \begin{equation}
        {\dot{c}}_1 = -\sum_{m'\mu'} \frac{d^2_{01}c\chi^2_{s'1}}{\hbar\epsilon_0 \pi b^4\alpha_{s1}} J^2_{m'}(\frac{\chi_{s'1}R}{b})\cos^2{(\frac{n'\pi z}{L})}\int_{c\chi_{s'1}/b}^{\infty} \dd \omega_{s'1} \int^t_0 \dd t' \frac{e^{i\left(\omega_{10}-\omega_{s^\prime 1}\right)(t-t')}}{\sqrt{\omega_{s'1}^2 - c^2\chi_{s'1}^2/b^2}} c_1(t').\label{explformofc1}
    \end{equation}
\end{widetext}
In the emission spectrum, the frequency of light associated with the emitted radiation is going to be centered about the eigen vibration frequency $\omega_{10}$. The quantity $\omega_{s'1}$ varies little around $\omega_{s'1}=\omega_{10}$ for which the time integral in Eq.(\ref{explformofc1}) is not negligible. We can then replace $\omega_{s'1}$ by $\omega_{10}$ and the lower limit in the $\omega_{s'1}$ integration by $-\infty$. Then we have
\begin{equation}
    \int_{-\infty}^{\infty} \dd \omega_{s'1} e^{i\left(\omega_{10}-\omega_{s^\prime 1}\right)(t-t')} = 2\pi\delta(t-t').
\end{equation}
Since the term $\sqrt{\omega_{s'1}^2 - c^2\chi_{s'1}^2/b^2}$ in the denominator has no physical meaning at less than 0, so that a step function will be added to Eq.(\ref{explformofc1}). The Eq.(\ref{explformofc1}) becomes
\begin{equation}
\begin{split}
    {\dot{c}}_1 =& \frac{3\varGamma^{(0)}_{10}}{\xi_{10}^3} \sum_{m'\mu'} \frac{\chi^2_{s'1}}{\alpha_{s'1}\sqrt{\xi_{10}^2 - \chi^2_{s'1}}} J^2_{m'}(\chi_{s'1}R/b)\\
    &\times\cos^2(\sqrt{\xi_{10}^2 - \chi^2_{s'1}}z/L) \theta(\xi_{10} - \chi_{s'1})c_1(t) ,\label{c1final}
\end{split}
\end{equation}
where $\xi_{10} = \omega_{10}a/c$, $\theta(x) = 1(x>0),0(x<0)$ and
\begin{equation}
    \varGamma^{(0)}_{10} = \frac{e^2 \abs*{\mel{\varPsi_0}{\vb*{D}}{\varPsi_1}}^2 \omega_{10}^3}{3\pi\hbar\varepsilon_0c^3}
\end{equation}
is the transition rate in free space. By using Eq.(4-12) in \cite{scully1999quantum}, we have
\begin{equation}
    {\dot{c}}_1 = -\frac{\varGamma_{10}}{2}c_1 ,
\end{equation}
where $\varGamma_{10}$ is the transition rate between $\ket{\varPsi_1}$ and $\ket{\varPsi_0}$ in cavity.
In this way the physical meaning of Eq.(\ref{c1final}) is clear: at $\xi_{10} < \chi_{011}$, transition between $\ket{\varPsi_1}$ and $\ket{\varPsi_0}$ is forbidden. Using the same approach similar to Weisskopf-Wigner approxiamation mentioned in \cite{scully1999quantum}, we can transform the summation on the right-hand side of Eq.(\ref{c0dot}) and  Eq.(\ref{c1dot}) to be expressed in terms of transition rates and get
\begin{align}
  {\dot{c}}_0&=-\frac{\varGamma_{21}}{2}c_0 ,\\
  {\dot{c}}_1&=-ig_\omega c_0e^{i\left(\omega_{21}-\omega_{s\sigma}\right)t}--\frac{\varGamma_{10}}{2}c_2 ,\\
  {\dot{c}}_2&=g_{\omega^\prime}c_2 e^{i\left(\omega_{10}-\omega_{s^\prime\sigma^\prime}\right)t} ,
\end{align}
where $\varGamma_{21}$ and $\varGamma_{10}$ are respectively

\begin{equation}
\begin{split}
    \varGamma_{21} = &\frac{6\varGamma^{(0)}_{21}}{\xi_{21}^3} \sum_{m\mu} \frac{\chi^2_{s1}}{\alpha_{s1}\sqrt{\xi_{21}^2 - \chi^2_{s1}}} J^2_m(\chi_{s1}R/b) \\
    &\times\cos^2(\sqrt{\xi_{21}^2 - \chi^2_{s1}}z/L) \theta(\xi_{21} - \chi_{s1}),\label{gamma21}
\end{split}
\end{equation}

\begin{equation}
\begin{split}
    \varGamma_{10} = &\frac{6\varGamma^{(0)}_{10}}{\xi_{10}^3} \sum_{m\mu} \frac{\chi^2_{s1}}{\alpha_{s1}\sqrt{\xi_{10}^2 - \chi^2_{s1}}} J^2_m(\chi_{s1}R/b)\\
    &\times\cos^2(\sqrt{\xi_{10}^2 - \chi^2_{s1}}z/L) \theta(\xi_{10} - \chi_{s1}).\label{gamma10}
\end{split}
\end{equation}

We are only interested in the system of $t\gg\varGamma^{-1}_{21},\varGamma^{-1}_{10}$, in this case $c_0\left(\infty\right)$ and $c_1\left(\infty\right)$ both tend to 0, and $c_2\left(\infty\right)$ is
\begin{widetext}
    \begin{equation}
      c_2(\infty)=\frac{g_\omega g_{\omega^\prime}}{\left[i\left(\omega_{s^\prime\sigma^\prime}+\omega_{s\sigma}-\omega_{20}\right)-\frac{\varGamma_{21}}{2}\right]\left[i\left(\omega_{s^\prime\sigma^\prime}-\omega_{10}\right)-\frac{\varGamma_{10}}{2}\right]}.\label{c3expl}
    \end{equation}
\end{widetext}

Subsituting Eq.(\ref{c3expl}) into Eq.(\ref{matterfockstate}), we obtain the biphoton state
\begin{widetext}
    \begin{equation}
    \ket{\Omega}=\sum_{s\sigma,s^\prime\sigma^\prime}\ \frac{-g_\omega g_{\omega^\prime}\left|\omega_{s\sigma},\omega_{s^\prime\sigma^\prime}\right\rangle}{\left[\left(\omega_{s^\prime\sigma^\prime}+\omega_{s\sigma}-\omega_{20}\right)+\frac{i\varGamma_{21}}{2}\right]\left[\left(\omega_{s^\prime\sigma^\prime}-\omega_{10}\right)+\frac{i\varGamma_{10}}{2}\right]}.\label{biphotonwavefunc}
\end{equation}
\end{widetext}

In order to obtain the specific biphoton wave function, we need to first obtain the transition rates between the second excited state and the first excited state and between the first excited state and the ground state. Then we use data from \cite{omegadata}. And subsequently, by using the relationship between $N$ and the energy levels
\begin{align}
  \omega_{10}=\frac{E_1-E_0}{\hbar}=\frac{w^2\hbar}{2m}\left(2N+3\right)=0.33 \space\text{eV} ,\notag\\
  \omega_{20}=\frac{E_2-E_0}{\hbar}=\frac{w^2\hbar}{m}\left(2N+2\right)=0.64 \space\text{eV} ,\\
  \omega_{21}=\frac{E_2-E_1}{\hbar}=\frac{w^2\hbar}{2m}\left(2N+1\right)=0.31 \space\text{eV} ,\notag
\end{align}
we can find $N=19.6$, which in turn allows us to find the coordinate operator matrix elements by using Eq.(\ref{matele}). Then the dipole moments are $d_{10}=0.16$ e\AA ~and $d_{21}=0.23$ e\AA.

Experimental results show that the radius of axons in the brain or CNS is generally 2$\sim$6 $\mu$m, and the thickness of myelin sheaths is generally 1$\sim$3 $\mu$m\cite{AxonPhysiology,ratio10078}. then let the radius of exposed axons (inner radius of myelinated neurons) be 2 $\mu$m, and the thickness ranges from 0 to 3 $\mu$m. Taking that the C-H bonds are widely distributed in between the inner and the outer {radii of the cylinder, we calculate the average of Bessel functions and $\cos$ functions in Eq.(\ref{gamma10}) and Eq.(\ref{gamma21})

\begin{equation}
    \int_{a}^{b} J^2_m(\chi_{s1}R/b) \dd R , \int_{0}^{L} \cos^2(\frac{n\pi z}{L}) \dd z.
\end{equation}

\begin{figure}[ht]
    \includegraphics[width = 8.6cm]{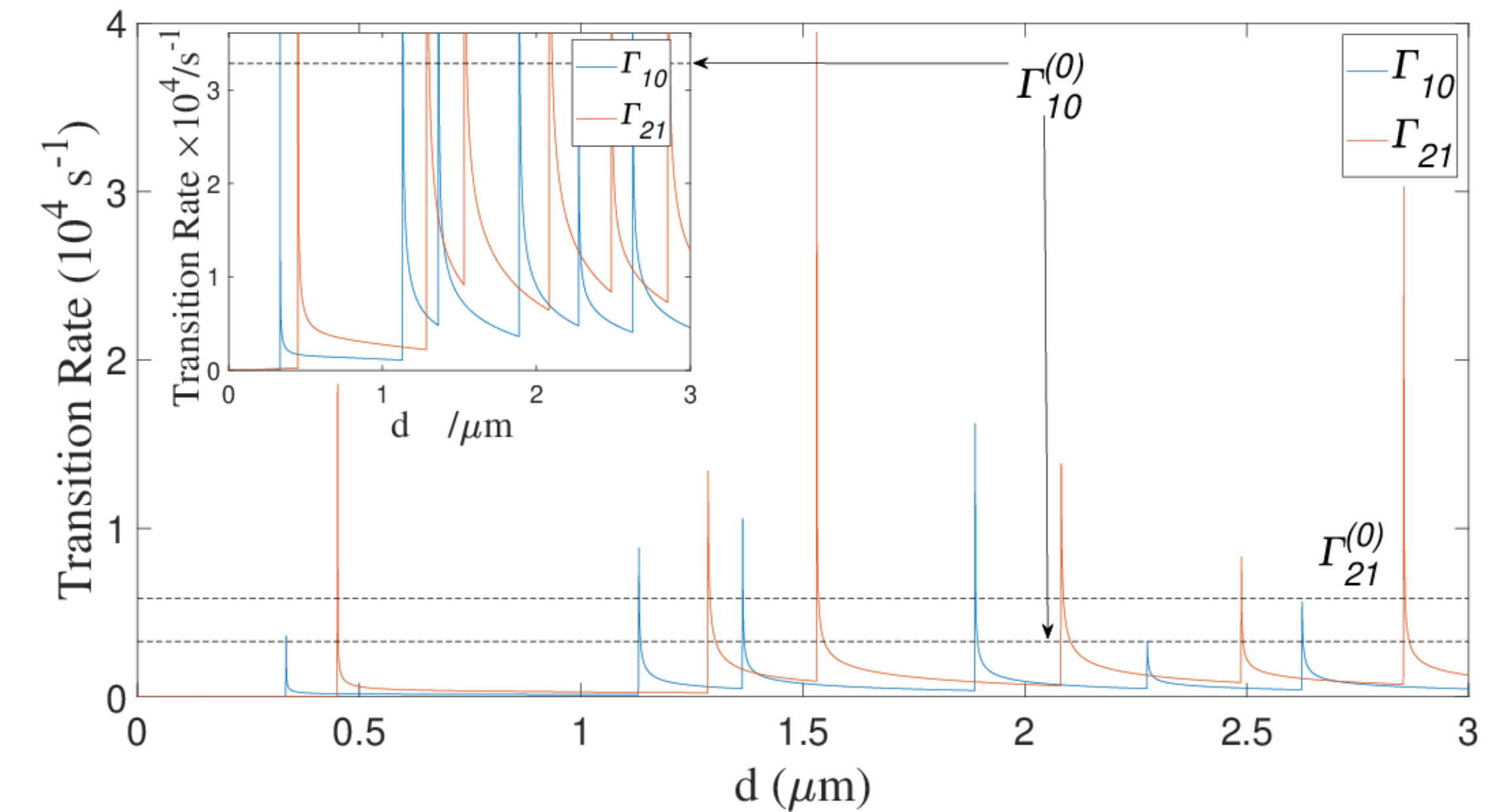}
    \caption{Transition rates $\varGamma_{21}$ and $\varGamma_{10}$. The sharp peaks correspond to having $\xi_{10}$ or $\xi_{21}$ equal to $\chi_{m\mu}$.}
    \label{fig:trans}
\end{figure}

Then we can observe the variation of the transition rates with myelin thickness as shown in Figure.\ref{fig:trans}. Each peak in the figure corresponds to a zero point of the Bessel function $\chi_{m\mu}$. When the cavity radius is less than 2.45 $\mu$m, $\varGamma_{21}$ and $\varGamma_{10}$ are quite small, which is due to the fact that the dipoles are all clustered at the axon boundaries $R/b \approx 1$, and so the coupling constant tends to 0. The coupling between the electromagnetic field and the vibrational modes is then negligible, and the vibrational modes change without any relationship between the excitation or non-excitation of the electromagnetic field.

\begin{figure}[ht]
    \includegraphics[width = 8.6cm]{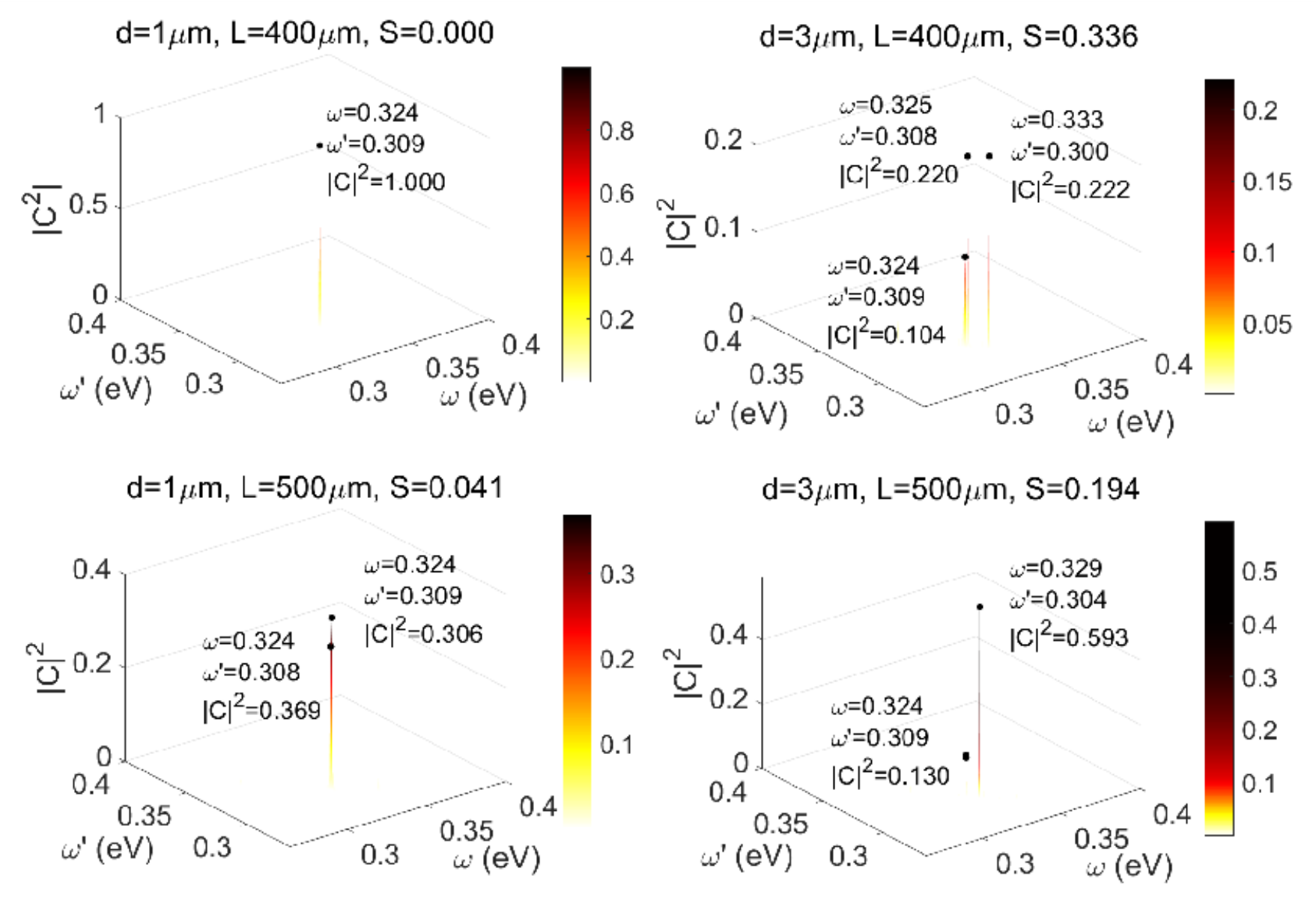}
    \caption{The square of amplitudes which are given by the coefficients in biphoton wave functions at different sizes of cavity. X-axis and Y-axis are in unit of electron volts (eV). The higher peaks in each case always map to the modes that close to $\omega_{10}$ and $\omega_{21}$. The values of von Neumann entropy are respectively 0.000, 0.336 in the first, and 0.041, 0.194 in the second row.}
    \label{fig:C}
\end{figure}

The biphoton wave function can be calculated using the transition rates and the coupling constants. Considering that the linewidth is much smaller than the vibrational frequency, the allowed modes are almost all at resonance. So, for practical calculations, we will disregard frequencies greater than 0.39 eV and less than 0.26 eV. From the coefficient plots (Figure.\ref{fig:C}) of the normalised biphoton wave functions it can be seen that the allowed modes are all very close to resonance or at resonance, and only a few modes are significantly allowed.

\section{evaluating entanglement through schmidt decompostion}\label{sec:schmidt}
In the previous section we obtained the wave function of biphoton system. We now evaluate the entanglement degree of this system. In quantum mechanics, the pure state of a bipartite system is described by a wave function $\ket{\Omega(x_1,x_2)}$, which $x_1$ and $x_2$ represent any degrees of freedom of the two particles: coordinates, momentum, polarization, or frequency, etc\cite{entanglementquanti}. In the discrete case, the wave function can be written as Eq.(\ref{biphotonwavefunc}). According to Schmidt analysis\cite{nielsen2010quantum}, it can always be expanded with two adjoint basis
\begin{equation}
    \ket{\Omega} = \sum_{n} \sqrt{\lambda_n} \ket{k_1}_n \ket{k_2}_n ,\label{schmidt}
\end{equation}
where $\ket{k_1}_n$ and $\ket{k_2}_n$ are two adjoint Schmidt modes. Representation of Eq.(\ref{schmidt}) is known as Schmidt decomposition. The expansion parameters $\sqrt{\lambda_n}$ in Eq.(\ref{schmidt}) are real and positive, and they obey the normalization condition $\sum_n \lambda_n = 1$. This decomposition is closely related to the singular value decomposition (SVD)\cite{schmidtmodes}. We can consider the coefficients in Eq.(\ref{biphotonwavefunc}) as a matrix $\bm{C}$, and then this matrix can always do SVD to obtain two unitary matrices $\bm{U}$ and $\bm{\Sigma}$ and a diagonal matrix $\Lambda$. The relation between them is
\begin{equation}
    \bm{C} = \bm{U\Lambda\Sigma}.
\end{equation}
The squares of the diagonal elements of matrix $\Lambda$ then correspond to the expansion parameters $\lambda_n$. The distinguishing feature of the Schmidt decomposition is that when a particle is measured to be in one of the Schmidt modes $\ket{k_1}_n$, then the other particle must be in the adjoint mode $\ket{k_2}_n$ rather than the other mode, and the probability of finding such a pair of particles is given by $\lambda_n$. When two particles are independent of each other, they are not in entanglement. At this point, their wave functions can simply be written as
\begin{equation}
    \ket{\Omega} = \ket{k_1}_n \ket{k_2}_n.
\end{equation}
We get only one pair of Schmidt modes corresponding to the Schmidt decomposition of Eq.(\ref{schmidt}). When two particles are entangled, the Schmidt decomposition has more than one term. We can use the expansion parameters in Eq.(\ref{schmidt}) to define the von Neumann entropy to evaluate the manitude of entanglement, and the von Neumann entropy is defined as\cite{nielsen2010quantum}
\begin{equation}
    S = -\sum_n \lambda_n \log_2(\lambda_n).
\end{equation}
When the Schmidt number has only one term, the entropy is 0, indicating that the system is not entangled. As defined, the maximum value of entropy is related to the number of states of the particle, and the maximum value of von Neumann entropy is $\log_2 n$ when the particle has $n$ states.For normalisation purposes, the entropy involved in what follows is calculated in such a way that the base of the logarithmic function is taken to be the number of states n of the corresponding wave function.

\begin{figure}[htbp]
    \centering
    \includegraphics[width = 8.6cm]{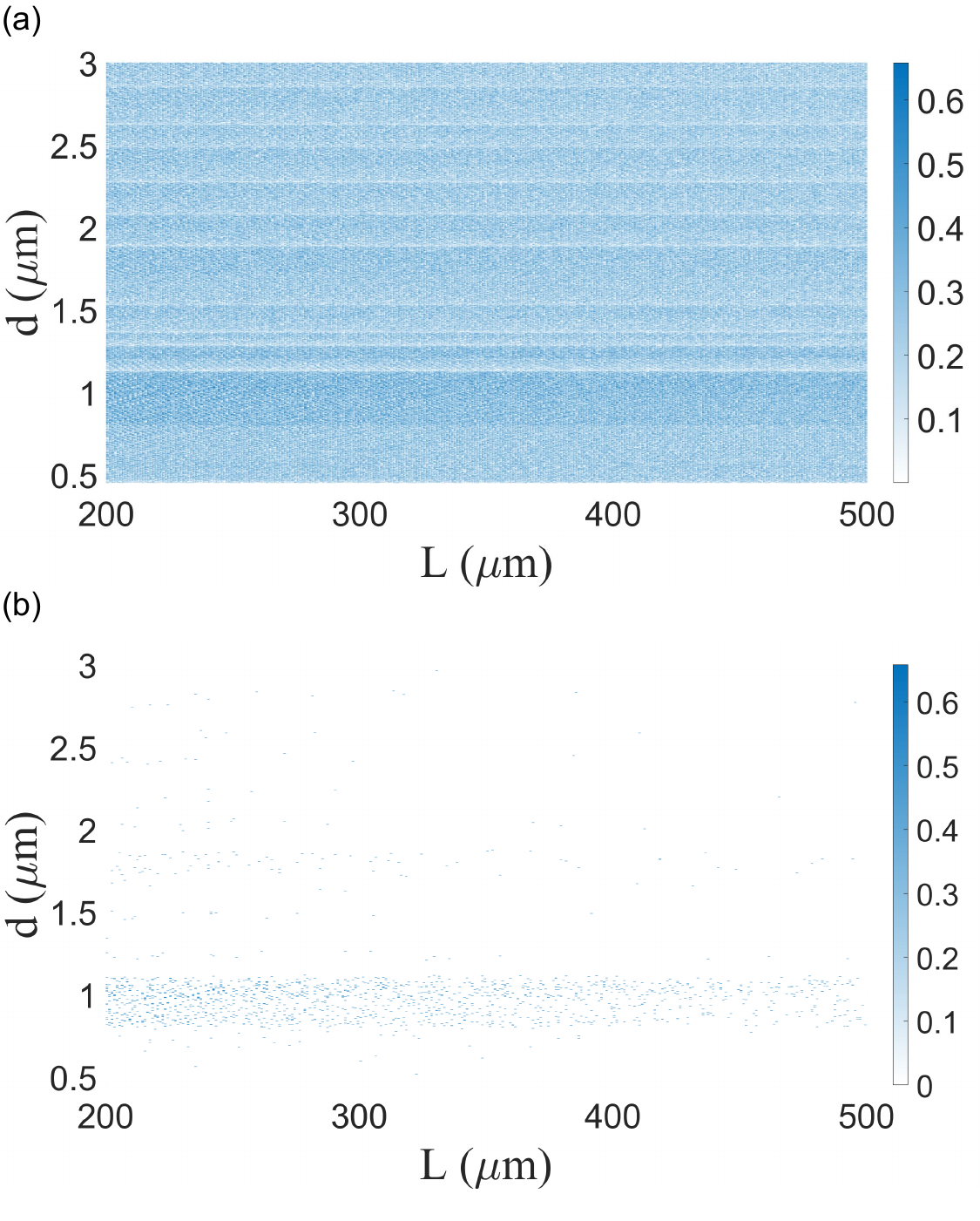}
    \caption{von Neumann entropy at different sizes of the cavity. (a) Colors from white to dark blue represent entropy values from 0 to 0.65. Myelin thicknesses ranged from 0.45 to 3 $\mu$m, and lengths ranged from 200 to 500 $\mu$m. (b) The relationship between myelin thickness and the degree of entanglement can be better seen by erasing the points in (a) with entropy values below 0.5. Entanglement is substantial at thicknesses between 0.8 and 1.1 $\mu$m, and it decays rapidly as the thickness becomes smaller.}
    \label{fig:S}
\end{figure}

The axon is usually wrapped by myelin sheath separated by several nodes of Raniver, each of which is usually 10$\sim$100 $\mu$m\cite{axonstruc,AxonPhysiology}. In the actual calculations, we neglect the total length of nodes of Raniver to be in the range of 200$\sim$500 $\mu$m, which affects the denseness of electromagnetic modes in the microcavity. When the thickness of the myelin sheath is less than 0.45 $\mu$m, the transition rates $\varGamma_{21}$ and $\varGamma_{10}$ are extremely small due to the very small coupling constant. We consider that no biphoton state will be generated when the thickness of the myelin sheath is in the range of 0$\sim$0.45 $\mu$m in the calculation. Electromagnetic modes at $(2,2,0)$ have energies exceeding the eigen vibrational frequency $\omega_{10}$, so modes exceeding $\chi_{221}$ are not considered here, and the results obtained are shown in Figure.\ref{fig:S}(a).

Owing to the discrete energy levels within the myelin cavity, instances where electromagnetic modes reside on both sides of the vibrational frequencies $\omega_{10}$ and $\omega_{21}$—and are in proximity—lead to significant entanglement. This arises from photons being indistinguishable particles, implying that measuring the frequency of one photon imparts partial information about the other. Conversely, when a discrete electromagnetic mode precisely corresponds to $\omega_{10}$ and $\omega_{21}$, it dominates, resulting in minimal entropy.  The molecular coupling constant in Eq.(\ref{couple}) are about an order of magnitude of 10$^7$, and the transition rates (natural line width) is about an order of magnitude of 10$^4$, which is small and negligible compared to the frequencies of the photon and vibrational energy levels in the denominator, which are about 10$^{14}$. The discrete energy levels are solely determined by myelin geometry, making the creation and availability of entanglement entirely contingent on myelin size. In contrast to cascade radiation of the vibrational spectrum in free space, characterized by extremely narrow linewidths and continuous electromagnetic modes, photon cascade radiation within the myelin sheath has only one mode—photons corresponding to frequencies $\omega_{10}$ and $\omega_{21}$.

\section{discussions}

To summarize, the results of the cascade photons emission process by cQED and quantum optics indicate that biphotons in quantum entanglement can be released through cascade radiation on the vibrational spectrum of C-H bonds in the tails of lipid molecules inside cylindrical cavities encased by neural myelin. The presence of discrete electromagnetic modes due to the cavity structure formed by the myelin sheath, distinguishing it from the free-space continuous electromagnetic modes, results in the frequent production of highly entangled photon pairs permitted within the myelin cavity. Notably, due to the presence of microcavities, the coupling can be significantly enhanced compared to that in free space, indicating a higher probability of emitting photons. It should be noted that our model is very crude. The actual electromagnetic field should take into account the coupling of photons to the vibron ensembles, i.e. polaritons, which should be considered in future studies.

As shown in Figure.\ref{fig:S}(b), the degree of entanglement is relatively high when the thickness of the myelin sheath is between 0.8$\sim$1.1 $\mu$m. Taking the radius of the axon as 2 $\mu$m, this corresponds to a ratio of the inner/outer radii 0.65$\sim$0.72, close to the literature value of 0.6$\sim$0.8\cite{ratio10078,axondiameter}.  The entanglement decreases rapidly as the myelin thickness decreases beyond this ratio. Clinical results show that myelin becomes thinner with age and the likelihood of neurodegenerative diseases increases\cite{agemyelin,ageplastic}. These observations may underline a further relationship between the two phenomena.

Finally,  as a nature of such research we may excise some speculations at this stage. It was experimentally shown that the mid-infrared light energy at 53.53 THz has a modulating effect on the activity of K$^+$ ion channels. Furthermore, the activity is strong and not proportional to the number of photons\cite{songpnas}. One explanation is that the eight C=O bonds in an ion channel may be in a critical state of maximum superradiation. Such state has the energy of three excited C=O bonds whose value equals roughly to that of photon energy released by a single C-H bond. As the energy scales match up, it is plausible that the entanglement of the photons can pass on to the ion channels. Namely, entangled photon pairs emitted through cascade radiation can link the K$^+$ ion channels at different positions by entanglement. When one channel is activated by neuron, it affects the state of the other ones via quantum measurement, creating likely non-local correlations among them.

Polaritons inside a myelin involves a large number of vibronic states\cite{songneuron}. As a result the effect of thermal fluctuations on the states of the polariton (i.e. photons) can be negligible. In recent years similar systems such as cold-atom ensembles\cite{quantummem} were used as quantum memories, which can protect quantum entanglement between photons. When the local entanglement generated in each neuron could spread over to a larger region through entanglement swapping\cite{RevModPhys.81.865}, the neurons in the brain becomes further correlated. In this way, the entanglement can propagate out within the neuromedullary sheaths,  serving as a quantum communication resource in the nervous system.  It can possibly offer a mechanism of over-distance synchronization.

\begin{acknowledgments}
Z. L thanks Dr. Y. Y. Wang for fruitful discussion. This work was supported in part by the National Natural Science Foundation of China (YCC, Grant Approval No. 12375034).
\end{acknowledgments}
%

\end{document}